\documentclass[twocolumn,superscriptaddress,showpacs,preprintnumbers,amsmath,amssymb,aps]{revtex4}

\usepackage{graphicx}
\usepackage{dcolumn}
\usepackage{bm}
\usepackage{longtable}

\begin{document}

\def\f{\frac}
\def\rdown{\rho_{\downarrow}}
\def\pa{\partial}
\def\th{\theta}
\def\Ga{\Gamma}
\def\ka{\kappa}
\def\bea{\begin{eqnarray}}
\def\eea{\end{eqnarray}}
\def\be{\begin{equation}}
\def\ee{\end{equation}}
\def\pa{\partial}
\def\d{\delta}
\def\K{\kappa}
\def\a{\alpha}
\def\eps{\epsilon}
\def\th{\theta}
\def\na{\nabla}
\def\nn{\nonumber}
\def\lan{\langle}
\def\ran{\rangle}
\def\pr{\prime}
\def\rarrow{\rightarrow} 
\def\larrow{\leftarrow}


\title{Hartman effect in presence of Aharanov Bohm flux}
\author{Swarnali Bandopadhyay}
\email{swarnali@bose.res.in}
\affiliation{S. N. Bose National Centre for Basic Sciences, JD Block,
Sector III, Salt Lake City, Kolkata 700098, India}
\author{Raishma Krishnan}
\email{raishma@iopb.res.in}
\author { A. M. Jayannavar }
\email{jayan@iopb.res.in}
\affiliation{Institute of Physics, Sachivalaya Marg, 
Bhubaneswar 751005, India }
\begin{abstract}
Abstract: 
The Hartman effect for the tunneling particle implies the independence of group
delay time on the opaque barrier width, with superluminal velocities as a
consequence. This effect is further examined on a quantum ring geometry in the 
presence of Aharonov-Bohm flux. We show that while tunneling through an opaque 
barrier the  group delay time for given incident energy becomes independent of 
the barrier thickness as well as 
the magnitude of the flux. The Hartman effect is thereby extended beyond one 
dimension and in the presence of Aharonov-Bohm flux.

\end{abstract}



 \maketitle
 \section{Introduction}
\label{s1}
The concept of tunneling in the realm of quantum mechanics has 
gained much attention over many years. The immense potentiality 
of this concept has led to the study of various 
time scales to understand a time the particle takes 
to tunnel through a barrier~\cite{landauer,hauge,anantha}. 
The group delay time associated with potential scattering is one of the 
important quantities related efficiently to the 
dynamical aspect of scattering
in quantum mechanics. The time delay for scattering 
processes can be calculated by following the peak of 
a wavepacket~\cite{wigner}. The phase delay time 
($\tau$) is expressed in terms of the derivative of the 
phase shift of the scattering matrix with respect to energy.
Since its inception, Wigner phase delay time has been a quantity of interest 
from fundamental as well as technological point of view. The delay time 
statistics is intimately connected with the dynamic admittance of 
microstructures~\cite{buttiker}. This delay time is also directly related
to the density of states~\cite{buttiker2}. The universality of the delay time 
distributions in random and chaotic systems has been established earlier 
~\cite{jayan}. 

The explicit calculation of group delay time in the problem of a particle 
tunneling through a rectangular barrier becomes independent of the barrier 
width in the case of an opaque barrier~\cite{hartman,fletcher}. 
This phenomenon, often referred as `Hartman effect', implies that for 
sufficiently large barriers the 
effective group velocity of the particle inside the barrier can become 
arbitrarily large ~\cite{olkhovsky}. In other words, the evanescent waves can 
travel with superluminal speeds. 

Though experiments with electrons for verifying 
this prediction is yet to be done, the formal identity between 
the Schr\"odinger equation and 
the Helmholtz equation for electromagnetic wave 
correlates the results for electromagnetic and microwaves 
to that of electrons. Photonic experiments show 
that electromagnetic pulses travel with group velocities in excess of the 
speed of light in vacuum as they tunnel through a constriction in a
waveguide ~\cite{nimtz}. Other experiments with photonic band-gap structures 
also verified that `tunneling photons' travel with superluminal group 
velocities ~\cite{steinberg}. Thus all these experiments directly or indirectly
confirmed the occurence of the Hartman effect without violating so called 
`Einstein causality'. Operationally, superluminal velocities 
have been measured in terms of delay time between the appearance 
of a pulse peak at the input and 
a pulse peak at the output of a barrier. It should also be noted that the 
method based on following the peak of the wavepacket looses its significance 
under strong distortion of wavepacket~\cite{landauer}. Moreover, there is no 
causal relationship between the peak of transmitted wavepacket and that of the
incident wavepacket. This is due to the fact that peak of the transmitted 
wavepacket can leave the scattering region long before the peak of the 
incident wavepacket has arrived. However, Hartman effect and its origin is 
still considered as a poorly resolved problem. Very recently 
Winful~\cite{winful2,winful3} argued that `the tunneling particle 
or wave packet is not really traveling with superluminal velocity but actually 
a standing wave, that just stand and waves!'. The incident wave simply 
modulates this standing wave. The output adiabatically follows the input 
with delay proportional to the stored energy. It is shown that the short time 
delay observed is due to energy storage and release and has nothing 
to do with real propagation and hence should not be linked with velocity. 
Thus the origin of the Hartman effect is traced to stored energy. Since 
the stored energy in the evanescent field decreases exponentially within 
the barrier after a certain decay distance it becomes independent of the 
width of the barrier. 

All the studies on Hartman effect till date are restricted to one dimension 
only. Mostly single~\cite{olkhovsky} or successive rectangular
barriers~\cite{recami} have been considered. A very simple semiclassical
derivation of Hartman effect valid for potential barrier of general shape
in one dimension has been presented in ref.~\cite{bianchi}. In our present 
work we verify Hartman effect beyond one dimension and in the presence of 
Aharonov-Bohm flux. 

\section{Description of the system}

\begin{figure}[hbp!]
 \begin{center}
\input{epsf}
\hskip15cm \epsfxsize=3in \epsfbox{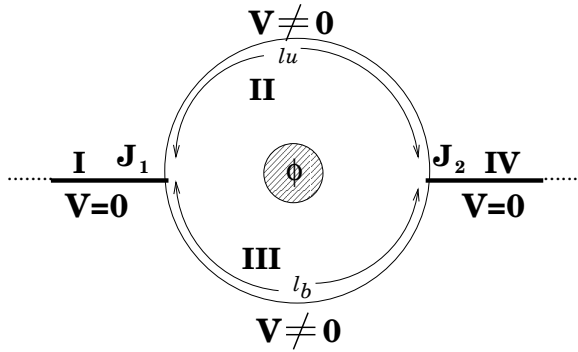}
 \caption{Schematic diagram of a ring connected to two 
leads in the presence of an Aharanov-Bohm flux.} \label{system}
 \end{center}
 \end{figure}

We study the scattering problem across a quantum ring 
geometry as schematized in Fig.~\ref{system}. Such ring geometry systems have 
been extensively investigated in mesoscopic physics in 
analysing normal state Aharanov-Bohm 
effect which has been observed experimentally~\cite{webb,gefen}. 
Our system of interest constitute a loop 
connected to two semi-infinite ideal wires in the presence of a magnetic 
flux as in Fig.~\ref{system}. There is a finite quantum mechanical potential 
$V$ inside the loop while that in the connecting 
leads are set to be zero. We focus on a situation wherein 
the incident electrons have an energy $E$ less than $V$. 
The impinging electrons in this subbarrier 
regime travels as an evanescent mode throughout 
the circumference of the loop and the transmission or the conductance 
involves contributions from both the Aharanov-Bohm effect as well as 
quantum tunneling. We are interested in a single channel case where the 
Fermi energy lies in the lowest subband. To excite the evanescent modes 
in the ring we have to make the 
width of the ring much less than that of the connecting wires. The electrons 
occupying the lowest subband in the connecting wire on entering the ring 
experience a higher barrier (due to higher quantum zero point energy)
and propagate in the loop as evanescent mode.
The transmission or conductance across such systems has been 
studied in detail~\cite{jayan3,gupta}. 
In this work an analysis of the phase delay time or 
the group delay is carried out.

\section{Theoretical treatment}

We approach this scattering problem using  the quantum wave guide theory 
~\cite{xia,jayan2}. The wave function 
in different region in absence of magnetic flux are given below 

\begin{eqnarray}
\psi_I(x_1) &=& e^{ikx_1} + r e^{-ikx_1}\, , \label{wv1}\\
\psi_{II}(x_2)&=& A\, e^{i\,q\,x_2} + B\, e^{-i\,q\,x_2}\, , \label{wv2}\\
\psi_{III}(x_3) &=& C\, e^{i\,q\,x_3} + D\, e^{-i\,q\,x_3} \, ,\label{wv3}\\
\psi_{IV}(x_4) &=& t\, e^{i\,k\,x_4} \, ,\label{wv4}
\end{eqnarray}
with $k=\sqrt{2mE/\hbar^2}$ being the wavevector of the incident 
propagating electrons in the leads and $q=\sqrt{2m(E-V)/\hbar^2}$ 
the wavevector in the ring. We have assumed the origin of the co-ordinates of 
$x_1$ and $x_2$ to be at $J_1$ and that for $x_3$ and $x_4$ to be at $J_2$. 
At $J_1$, $x_3=l_b$ and at $J_2$, $x_2=l_u$ where $l_u$ and $l_b$ are the
lengths of the upper and lower arms of the ring. Total circumference of the 
ring is $L=l_u + l_b$.

We use the Griffith boundary conditions~\cite{griffith}
\begin{equation}
\psi_{I}(0)=\psi_{II}(0)=\psi_{III}(l_b) \, ,\label{bc1}
\end{equation}
and 
\begin{equation}
\Sigma_i \frac{\partial \psi_i}{\partial x_i} =0 \, ,\label{bc2}
\end{equation}
at the junction $J_1$. All the derivatives are either outward or inwards 
from the junction~\cite{xia}. Similar boundary conditions hold for $J_2$ 
as well. We choose a gauge for the vector potential in which the 
magnetic field  appears only in the boundary conditions rather than 
explicitly in the Hamiltonian~\cite{xia,gefen}. Thus the electrons 
propagating clockwise and anticlockwise will pick up opposite phases. 
The electrons propagating in the clockwise direction 
in the upper arm from $J_1$ will pick up a phase $i\,\a$ at $J_2$ 
and electrons propagating anticlockwise from $J_2$ to $J_1$ 
in the upper arm pick up a phase $-i\,\a$ at $J_1$. 
Similarly, an electron picks up a phase $i\,\beta$ at 
$J_1$ moving in the clockwise direction from $J_2$ in the lower arm and 
$-i\,\beta$ at $J_2$ moving anticlockwise from $J_1$ 
in the lower arm of the loop. The total phase around the loop is 
$\a\,+\,\beta\,=\,2\,\pi\,\phi/\phi_0$, where $\phi$ and 
$\phi_0$ are the magnetic flux and flux quantum, respectively. 
Hence from above mentioned boundary conditions we get for propagating
waves~\cite{jayan3,jayan2,gupta,deo} 
\begin{eqnarray}
1+r\,&=&\, A + B\,e^{-i\,\alpha} \nonumber \\ 
&=&C\, e^{i\,q\,l_b}\,e^{i\,\beta}+ D\,e^{-i\,q\,l_b}\, ,
\label{bcwvJ1}\\ 
t&=&A\, e^{i\,q\,l_u}\,e^{i\,\alpha}+ B\,e^{-i\,q\,l_u} \nonumber \\
 &=& C\,+\,D\,e^{-i\,\beta}\, ,
\label{bcwvJ2}
\end{eqnarray}
\begin{eqnarray}
i\,k\,(1-r) &=&\,i\,q\,A\,-\,i\,q\,B\,e^{-i\,\alpha}\nonumber \\  
&-&i\,q\,C\,e^{{i\,q\,l_b}}\,e^{i\,\beta}\,+
\,i\,q\,D\,e^{-i\,q\,l_b}\, ,
\label{bccurrJ1}\\
i\,k\,t&=&i\,q\,A\,e^{i\,q\,l_u}\,e^{i\,\alpha}-i\,q\,B\,e^{-i\,q\,l_u}\,
\nonumber \\ 
&-&\,i\,q\,C\,+\,i\,q\,D\,e^{-i\,\beta}\, .
\label{bccurrJ2}
\end{eqnarray}
 \section{Results and Discussions}
\begin{figure}[htp!]
\includegraphics [width=7.5cm]{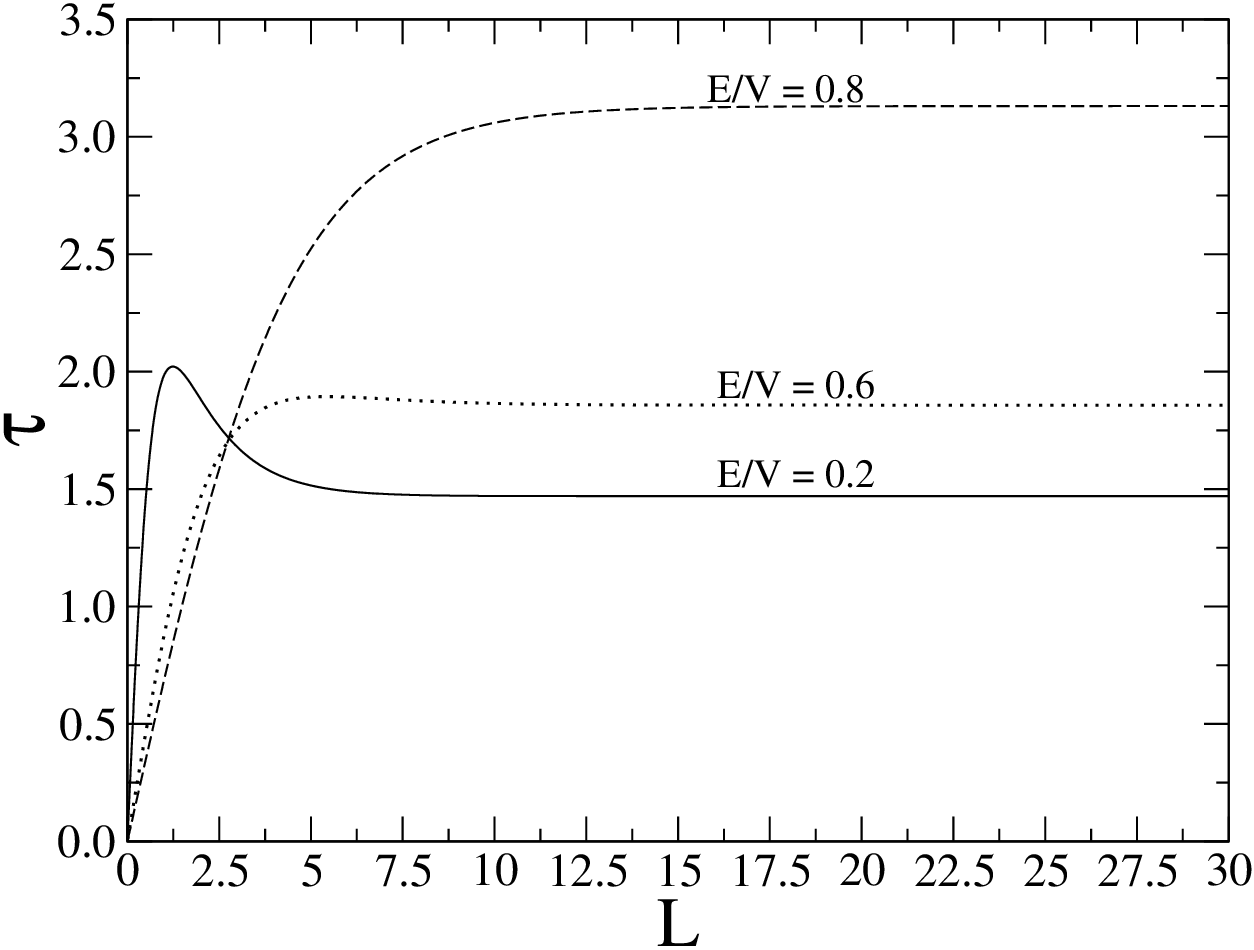}
\caption{Plot of $\tau$ versus $L$ for three different values 
of $E/V$ with $\phi=0$ and 
 $l_u=l_d$.  \label{tauvsL}}
\vspace{1.1cm}
\includegraphics [width=7.5cm]{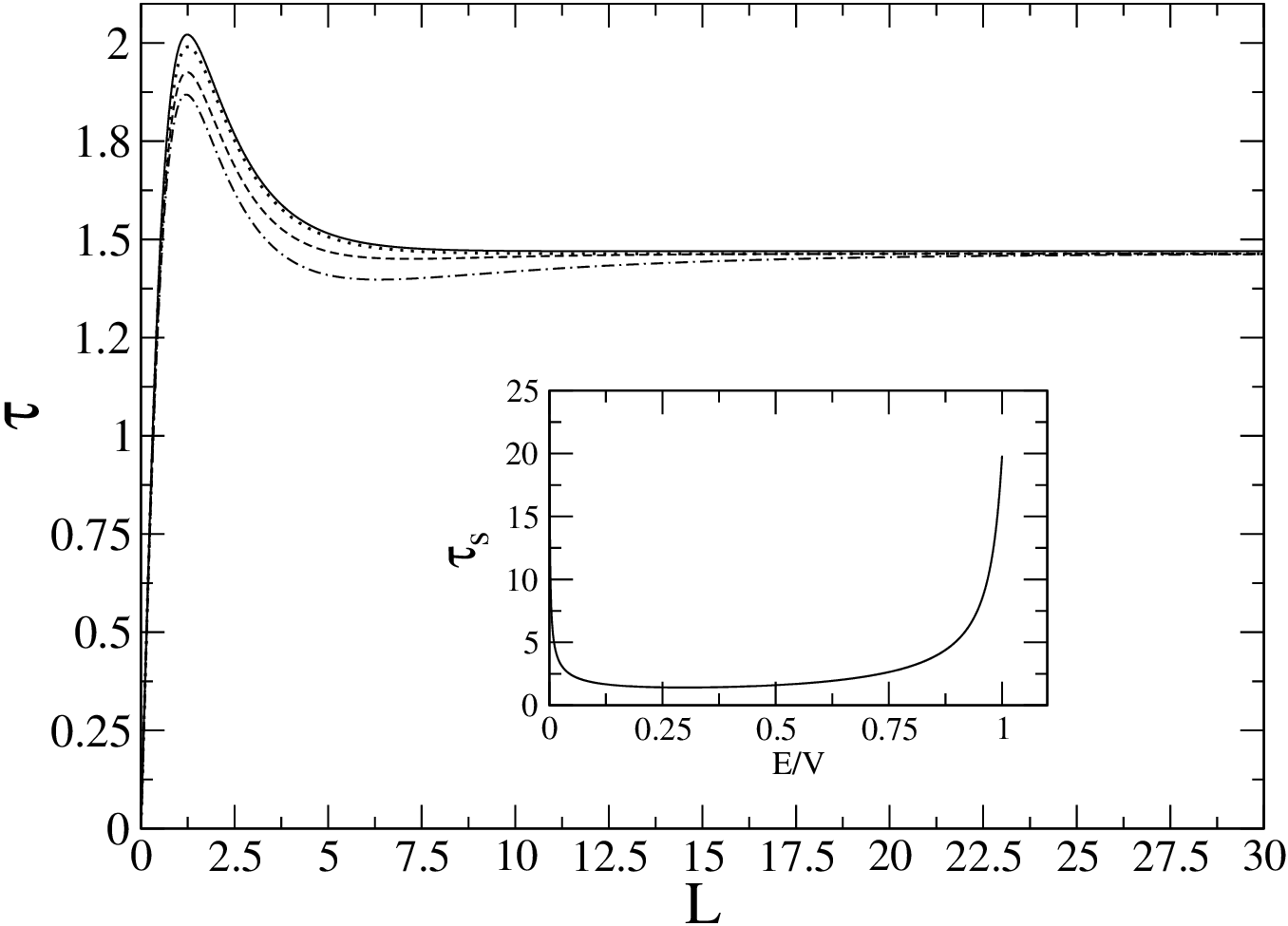}
\caption{Plot of $\tau$ versus $L$ 
for different arm length ratios. 
The ratio $l_u:l_b$ for the solid, dotted, dashed and 
dot-dashed curves are $1:1$, $3:2$, $4:1$, $9:1$ respectively.
Inset shows $\tau_s$ versus $E/V$.
\label{tauvsdiffLE}}
\vskip1.1cm
\includegraphics [width=7.5cm]{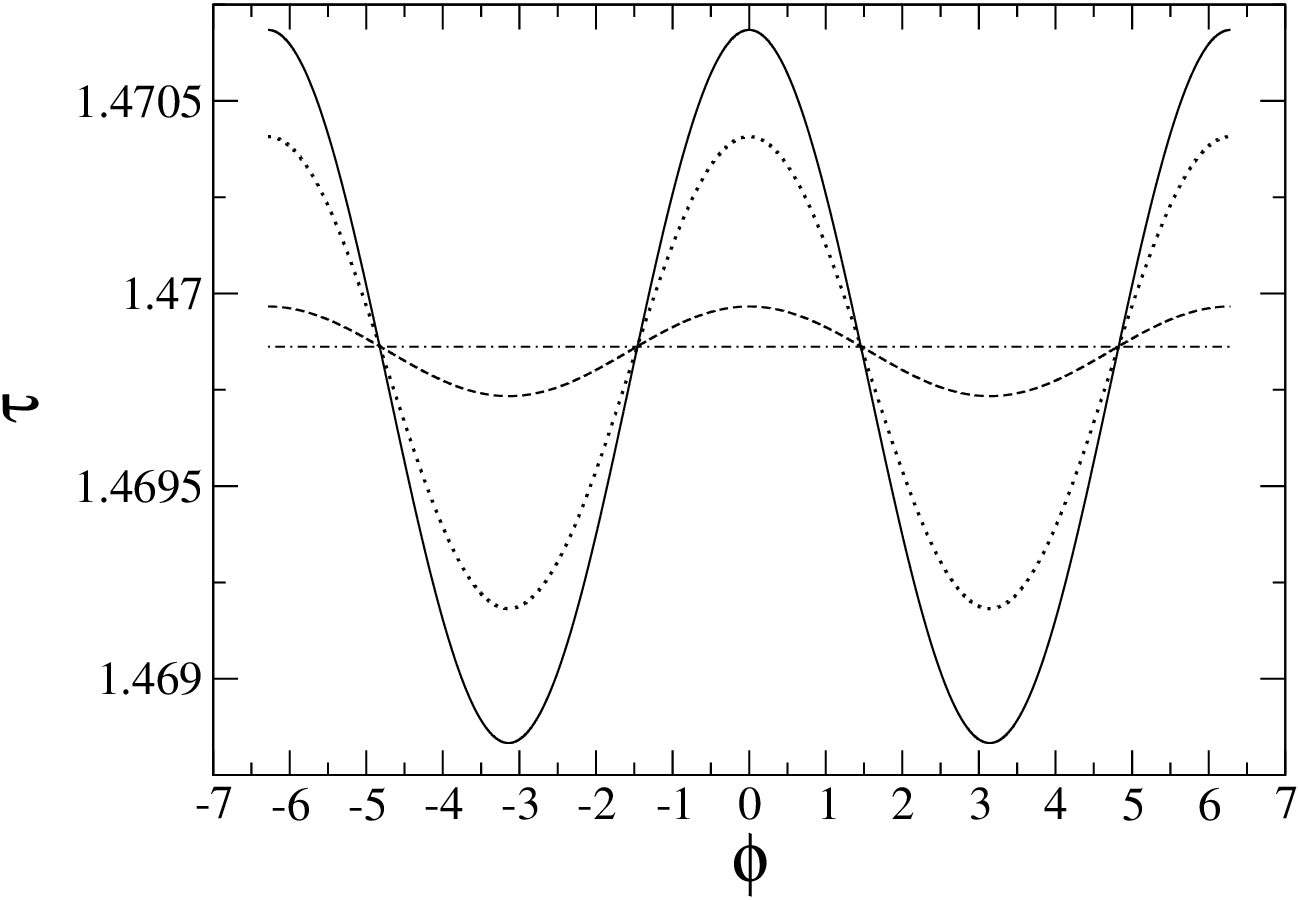}
\caption{Plot of $\tau$ versus $\phi$
for different $L$.
The solid, dotted, dashed and dot-dashed curves are for $L=10,\,10.5,\, 
12.5,\, 30 $ respectively. \label{tauvsphi}}
\end{figure}

For the evanescent regime in our problem we replace the wavevector 
$q$ in the loop 
by $i\,\kappa$ ($\kappa = \sqrt{2m(V-E)/\hbar^2}$). Solving Eqns.~(\ref{bcwvJ1})
-~(\ref{bccurrJ2}) we obtain an analytical expression 
for the transmission coefficient $t$ as
\be
t=\frac{4ik\K e^{i\a} \big[P e^{\K l_u}
+ Q e^{\K l_b}  \big]}
{PQk^2 + 2ik\K S_{-} 
+ 4 \K ^2 \big[e^{\K (l_u + l_b)}\big(e^{2i(\a+\beta)}+1\big) - S_{+}\big]}\label{texpr}
\ee
where 
\bea
P &=& e^{i(\a+\beta)}\big(e^{2\K l_b} - 1\big)\, ,\nn\\
Q&=&\big(e^{2\K l_u} -1\big)\, ,\nn\\
S_{\pm}&=&e^{i(\a+\beta)}\big(e^{2\K (l_u+l_b)} \pm 1\big)\,.\nn
\eea

 Next we evaluate the group delay time which is given by the 
 energy derivative of the phase of the transmission coefficient
 \begin{equation}
 \tau = \hbar \frac{\partial arg[t]}{\partial E}.
 \label{phtm}
 \end{equation}
We have set units of $\hbar$ and $2m$ to be unity. All the 
physical quantities are taken in dimensionless units 
($E\equiv E/V$,\,$\tau\equiv V\tau$
and $L \equiv L\sqrt{V}$). In Fig.~(\ref{tauvsL}) we plot phase time 
$\tau$ as a function of 
length $L$ of the ring for different values of incident energies 
in the absence of magnetic flux $\phi$ for the case where two armlengths $l_u$
and $l_b$ are equal. From the figure 
we clearly see that $\tau$ evolves as a function of length $L$ and 
asymptotically saturates to a value ($\tau_s$) which is independent of 
$L$ thus confirming
the Hartman effect. The saturation value increases with increasing incident
energy and the corresponding values for $E=0.2V$, $0.6V$ and $0.8V$ are 
$1.47,\, 1.86\,$ and $3.13$ respectively.
In one dimensional single barrier case in the tunneling region $\tau$
saturates to a constant value as a monotonic function of length. In our present case we observe depending on energy it is a monotonic or nonmonotonic function
as seen from Fig.~(\ref{tauvsL}). Our system differs from the one dimensional 
case in such a way that electron entering the ring can traverse along different alternative paths before transmitting. The interference between these 
alternative paths is responsible for this behaviour in the small $L$ regime 
where contributions from both evanescent and anti-evanescent modes dominate.
In Fig.~(\ref{tauvsdiffLE}) we plot the phase time versus $L$ for 
a particular energy, $E=0.2$, in the absence of magnetic flux $\phi$
 but for different length ratios of the upper and lower arms. 
We observe that the saturation value of the phase time is 
independent of the arm length ratios for a given 
energy as one can anticipate.
In the inset of Fig.~(\ref{tauvsdiffLE}) we plot $\tau_s$ versus $E/V$ for 
$\phi=0$, $L=30$ with equal upper and lower arm lengths. 
Plots with different armlength ratios ($l_u:l_b$) with different $\phi$ 
in the asymptotic limit were
found to overlap with the above curve in the entire energy regime. Moreover, 
it is
given by an analytical expression, $\tau_s\,=\,
(4\,\K^3\,+\,5\,k^2\,\K\,+\,(k^4/\K))/(2\,k\,((2\,\K^2\,-\,(k^2/2))^2+4\, 
k^2 \,\K^2))$ which agrees perfectly well with the numerical results.

In Fig.~(\ref{tauvsphi}) we have plotted group delay time as 
a function of flux $\phi$ for various values of circumference of the ring. 
In this case $l_u=l_b$ and $E=0.2V$.
We observe that $\tau$ is flux periodic with periodicity $\phi_0$. This
is consistent with the fact that all the physical properties in presence of 
Aharonov-Bohm flux across the ring must be periodic function of flux with a 
period $\phi_0$~\cite{buttiker2,webb,datta-bk}. However, we observe that as 
we increase the length of the ring the visibility or the magnitude of 
Aharonov-Bohm oscillations decreases. Consequently in the large length limit
the visibility of Aharonov-Bohm oscillations vanishes as can be seen from
Fig.~(\ref{tauvsphi}). The constant value of $\tau$ thus obtained in the 
presence of flux is
identical to $\tau_s$ in the absence of the flux 
(see Fig.~(\ref{tauvsL})) in the large length regime. This result clearly 
indicates that the delay time in the presence of opaque barrier becomes not only independent of length of the circumference but also it is independent of the Aharonov-Bohm flux thereby observing the Hartman effect in the presence of Aharonov-Bohm flux. We also find that the behaviour of reflection delay time is 
same as transmission delay time as anticipated from general symmetry laws from 
the simple geometric structure considered in the present case.

\section{Conclusions}
We have verified the Hartman effect in a quantum ring geometry in the presence of Aharanov-Bohm flux. Our studies show that the group delay time for a given incident energy becomes independent of the barrier thickness as well as the magnitude of the flux for the case of opaque barrier. We have also obtained similar results on different geometric structures by including a potential well between two adjacent barriers. These results which will be reported elsewhere~\cite{swarnali} were found to agree with the interesting observations made in 
Ref.~\cite{recami}.  

\section{Acknowledgements}
One of us (SB) thanks Institute of Physics, Bhubaneswar for the local 
hospitality, where the present work was carried out.

 \end{document}